\documentclass[preprint,aps,showpacs]{revtex4}
\usepackage{amsmath}
\usepackage{graphicx}

\newcommand{\f}{\begin{equation}}
\newcommand{\ff}{\end{equation}}
\newcommand{\fa}{\begin{eqnarray}}
\newcommand{\ffa}{\end{eqnarray}}
\newcommand{\pia}{\pi_{an}}
\newcommand{\pit}{\pi_{Tn}}
\newcommand{\pip}{\pi_{\phi n}}
\begin{document}
\title{ Discrete gravity and its continuum limit }
\author{Yi Ling${}^{1,2}$}
\email{yling@itp.ac.cn}
\affiliation{%
${}^1$ CCAST (World Laboratory), P.O. Box 8730, Beijing
   100080, China}
\affiliation{%
${}^2$ Institute of Theoretical Physics,
 Chinese Academy of Sciences,
 P.O.Box 2735, Beijing 100080, China}

\begin{abstract}

Recently Gambini and Pullin proposed a new consistent discrete
approach to quantum gravity and applied it to cosmological models.
One remarkable result of this approach is that the cosmological
singularity can be avoided in a general fashion. However, whether
the continuum limit of such discretized theories exists is model
dependent. In the case of massless scalar field coupled to gravity
with $\Lambda=0$, the continuum limit can only be achieved by fine
tuning the recurrence constant. We regard this failure as the
implication that cosmological constant should vary with time. For
this reason we replace the massless scalar field by Chaplygin gas
which may contribute an effective cosmological constant term with
the evolution of the universe. It turns out that the continuum
limit can be reached in this case indeed.
\end{abstract}

\keywords{Discrete gravity, classical limit, Chaplygin gas.}

\pacs{04.60.Nc, 04.20.Dw, 04.60.Kz}
\maketitle
\section{Introduction}

The cosmological singularity is a longstanding and fundamental
problem in quantum cosmology. Nowadays both string theory and loop
quantum gravity have been taking this issue into account
seriously, see\cite{Khoury,Liu} and \cite{Bojowald} for recent
progress along both approaches respectively. Motivated from string
theory, the cosmological solution from big crunch to big bang has
been proposed in \cite{Khoury}, which can also be viewed as the
orbifold compactification of solutions in higher dimensional field
theory. This solution has very peculiar features. For instance it
can be viewed as the Milne universe but identified by the boost
such that it contains closed timelike loops and more badly, not
Hausdorff. On the other hand this solution plays a special role in
string/M theory since such spacetime is flat everywhere except at
the singularity and the scalar field keeps staying in the weak
coupling region while approaching to the singularity. Consequently
the behavior of strings nearby such singularity in context of
string/M theory has been studied. Although some positive results
have been obtained in \cite{Liu}, implying strings may pass
through such singularity, the instability of this solution was
also pointed out in \cite{Horowitz}. Therefore further delicate
consideration is needed to the understanding of cosmological
singularity in string theory.

Recently another novel program called as discrete gravity has shed
new light on this problem\cite{Gambini1} as well. The idea is to
discretize Einstein equations. But in contrast to usual lattice
quantum gravity with fixed multipliers which breaks diffeomorphism
invariance manifestly, in discrete gravity the lagrangian
multipliers have to vary with ``time'' level $n$ such that
discretized equations can be preserved consistent. Furthermore,
unlike ordinary mechanism of canonical quantum gravity, discrete
gravity contains an attractive feature of constraint free such
that many conceptual problems like the problem of time might be
solved in a more promising way\cite{Gambini2}. In context of
cosmology this program has also appealing advantages of evolving
through the Big-Bang singularity in a general fashion and
naturally introducing a notion of time at quantum mechanical
level.

In this paper we apply the discrete techniques to study the
cosmological model which describes the universe may evolve from a
big crunch to a big bang. In this context we will see our results
support the argument that the cosmological singularity can be
avoided in a general fashion. But with a distinct starting point
from \cite{Gambini1}, we prefer to considering the fate of
singularities in universes with $\Lambda=0$ because in the case of
$\Lambda\neq 0$, the classical singularities in vacuum are often
coordinate singularities in the absence of matter, although this
claim does not mean that the singularity at quantum mechanical
level must not genuine. The discrete theory of FRW cosmology with
$\Lambda=0$ has been considered in\cite{Gambini1}.  However, as
its authors pointed out, the discrete theory describing the
universe with $\Lambda=0$ often suffers from the absence of the
continuum limit, which has negative implications to the
quantization of the theory. We will also meet this trouble in next
section when the model with massless scalar field is considered.
One purpose of our paper is to see whether we may improve the
continuum behavior of the discrete theory with zero cosmological
constant. As a way coming out, we consider the universe dominated
by Chaplygin gas rather than the ordinary massless scalar field in
section four. Such an exotic fluid behaves as matter dust at early
time, but contribute a nonzero effective cosmological constant at
later time.

We organize the paper as follows. In next section we briefly
review the framework of discrete mechanics focusing on constrained
system. Then we turn to consider the model with a massless scalar
field with $\Lambda=0$ in section three. After discussing the
relations between solutions in discretized theory and those in
continuum field theory, we argue that the singularity may be
avoided in discrete cosmology. In section four we study Chaplygin
gas dominated universe to obtain a better continuum limit of such
universes with vanishing cosmological constant at early time.

\section{Discrete theory of constrained system}
In this section we briefly review the classical framework of
discrete theory of constrained system which we will use in next
sections. For more details and in particular the quantization of
discrete system we refer to \cite{Gambini1,Gambini2}. We start
with the action  of a continuum mechanical system\f S=\int dt
L_c(q,\dot{q}).\ff The discretized formalism is implemented by
splitting time into equal intervals $t_n=n\epsilon$ where
$\epsilon$ is an infinitesimal constant. As a result, all the
derivatives with respect to time are replaced by discrete first
order finite difference, \f q=q_n,\quad \dot{q}={q_{n+1}-q_n\over
\epsilon}.\ff The time integral in the action then is replaced by
a sum $S=\sum_{n=1}^{N}L(q_n,q_{n+1})$, where $L(q_n,q_{n+1})$ is
understood as the discretized Lagrangian at instant $n$ with
relations $L(q_n,q_{n+1})=\epsilon L_c(q,\dot{q})$.

For a constrained system, we can straightforwardly obtain
$L(q_n,q_{n+1})$ from $L_c(q,\dot{q})$ with the form, \fa
L(q_n,q_{n+1})&\equiv & L(n,n+1)\nonumber\\
&=& p_n(q_{n+1}-q_n)-\epsilon H(q_n,p_n)-\lambda_n
C(q_n,p_n),\label{dc}\ffa where $H$ is the discrete version of
Hamiltonian and $C$ denotes the constraint. Note in the last term
the parameter $\epsilon$ has been absorbed into $\lambda_n$. In
this form we have also defined the canonically conjugate momenta
as \f p_{n+1}={\partial L(q_n,q_{n+1})\over
\partial q_{n+1}}.\label{cm2}\ff

In terms of canonical conjugate pairs the equations of motion can
be written as, \fa P^q_{n+1}-P^q_n &=& -\epsilon {\partial
H(q_n,P^q_{n+1})\over
\partial q_n}-\lambda_n
{\partial C(q_n,P^q_{n+1})\over \partial q_n}\nonumber\\
q_{n+1}-q_n &=& \epsilon {\partial H(q_n,P^q_{n+1})\over \partial
P^q_{n+1}}+\lambda_n {\partial
C(q_n,P^q_{n+1})\over \partial P^q_{n+1}}\nonumber\\
C(q_n,P^q_{n+1}) &=& 0.\label{eqs}\ffa One significant difference
between these equations and those continuum ones is that in
discrete theory the lagrangian multipliers have to be fixed at
each instant in order for the discrete version of the
``constraints'' to be preserved.

Therefore given a continuum theory with constraints we may obtain
its discrete Lagrangian and equations of motion. Going further
along this approach, we have two important issues to consider. One
is to investigate whether the discrete theory may give rise to any
different prediction comparing with the continuum theory. The
other one is to check whether the continuum theory can be
recovered by taking the continuum limit of the discrete theory. In
next sections we will apply this framework to cosmology and see
these cosmological models may help us to clarify these questions.

\section{Big crunch/bang universe}

Consider a general 4d theory of a scalar field coupled to gravity.
The Lagrangian has the form \f S=\int_M(R\sqrt{-g}-{1\over
2}g^{\mu\nu}\partial_{\mu} \phi\partial_{\nu}\phi-V(\phi)),\ff
where we use the $(-, +, +, +)$ signature for spacetime. In
context of cosmology we adopt minisuperspace description. The
metric reduces in conformal coordinate system to \f
ds^2=a^2(t)\left[ -N^2(t)dt^2+dx_i^2\right],\ff where $a$ is the
scale factor and $N$ lapse function. We strict our attention in
four dimensional spacetime with vanishing $V(\phi)$ in this
section such that the lagrangian has the form \f {\cal L}={-1\over
N}(4\dot{a}^2-{1\over 3}a^2\dot{\phi}^2).\label{cl} \ff The
cosmological solution from big crunch to big bang has the form
\cite{Khoury}, \f a(t)=a(0)|t|^{1\over 2}\quad\quad\quad
\phi(t)=\phi(0)+\sqrt{3}ln|t|,\label{cs}\ff where $t$ is from
$-\infty$ to $+\infty$. Obviously this solution has a singularity
at $a=0$ and $\phi=-\infty$ as $t$ goes to zero. Such a
singularity implies that the lagrangian breaks down near the
neighborhood of that point. As in the quantum or semiclassical
theory of hydrogen whose apparently continued spectrum is actually
discrete at microscopic level, we propose the similar
consideration should be applicable to the semiclassical theory of
spacetime geometry.\footnote{This proposal has been strongly
supported by loop quantum gravity, where the spectra of the area
and volume operator are obtained at least at kinematical level and
always taking discrete values.} As a result we conjecture  a
discretized version of lagrangian (\ref{cl}) probably is a good
starting point to better understand the microscopic origin of the
universe. The classical solution (\ref{cs}) may be only valid
macroscopically, but should be achieved by taking continuum limit
of solutions in the discretized theory.

First consider the canonical formalism of the theory. We define
conjugate momenta to scalar factor $a(t)$ and $\phi(t)$, \f
\pi_a={\partial {\cal L}\over
\partial \dot{a}}={-8\dot{a}\over
N},\;\;\;\;\;\;\;\pi_{\phi}={\partial {\cal L}\over \partial
\dot{\phi}}={2a^2\dot{\phi}\over 3N},\ff while the conjugate
momentum of lapse function $N$, $\pi_N$, is zero. The hamiltonian
can be constructed as \f {\cal
H}=\pi_a\dot{a}+\pi_{\phi}\dot{\phi}-{\cal L}={-N\over
4}({{\pi_a}^2\over 4}-{3{\pi_{\phi}}^2\over a^2})\equiv NH.\ff The
discrete version of an action is obtained from (\ref{dc}) \fa
L(n,n+1)&=&\pip(\phi_{n+1}-\phi_n)+\pia(a_{n+1}-a_n)\nonumber\\
&+& {N_n\over 4}({\pia^2\over4}-{3\pip^2\over a_n^2}),\ffa where
$\epsilon$ has been absorbed into the multiplier $N_n$.

From (\ref{cm2}) one find the momenta at instants $n$ and $n+1$
respectively are \fa P^{\phi}_{n+1}&=&\pip\;\;\;\;\;\;\;
P^{\phi}_n=\pip\nonumber\\
P^{\pi_{\phi}}_{n+1}&=&0\;\;\;\;\;\;\quad
P^{\pi_{\phi}}_n=-(\phi_{n+1}-\phi_n)+{3N_n\pip\over 2a_n^2}\nonumber\\
P^{a}_{n+1}&=&\pia \;\;\;\;\;\;\; P^{a}_n=\pia-{3N_n\pip^2\over 2a_n^3}\nonumber\\
P^{\pi_{a}}_{n+1}&=&0\;\;\;\;\;\;\;\quad P^{\pi_a}_n=-(a_{n+1}-a_n)-{N_n\pia\over 8}\nonumber\\
P^N_{n+1}&=& 0\;\;\;\;\;\;\;\quad P^N_n={1\over 4}({\pia^2\over
4}-{3\pip^2\over a_n^2}).\ffa Taking $(a_n, \phi_n, P^{a}_{n+1},
P^{\phi}_{n+1})$ as the fundamental variables yields the following
compact equations
\fa P^{\phi}_{n+1}-P^{\phi}_n &=& 0\nonumber\\
P^{a}_{n+1}-P^{a}_n &=& {3N_n\pip^2\over 2a_n^3}\nonumber\\
\phi_{n+1}-\phi_n &=& {3N_n\pip\over 2a_n^2}\nonumber\\
a_{n+1}-a_n &=& {-N_n\pia\over 8}\nonumber\\
{(P^{a}_{n+1})^2\over 4 } &=& {3(P^{\phi}_{n+1})^2\over
a_n^2}.\label{eom}\ffa We see $P_n^\phi$ is a conserved quantity
independent of level $n$. For convenience we may set it as a
positive constant $C$. Plugging it into the last equation, we have
\f P^a_{n+1}=\pm {2\sqrt{3}C\over a_n}.\label{pa}\ff Then from the
other equations in (\ref{eom}) it's straightforward to obtain the
recurrence relation of the scale factor as
 \f {a_{n+1}\over
a_n}={a_n\over a_{n-1}}\equiv q,\label{rr}\ff where $q$ can be any
positive constant.  Given a fixed $q$ and initial conditions, we
have two kinds of solutions corresponding to the different signs
of $P^a_{n+1}$ in (\ref{pa}).
\fa a_n &\sim & a_0q^n,
\;\;\;\;\;\;\;\;\;\;\quad\quad\quad \phi_{n} \sim
\phi_0+2\sqrt{3}n(q-1)\nonumber\\N_n  & \sim &
{4a_0^2q^{2n}(q-1)\over \sqrt{3}C}, \;\;\;\;\; P^a_{n+1}  \sim
-2\sqrt{3}{C\over a_n}\label{s1}\ffa or \fa a_n &\sim &
a'_0q'^n,\;\;\;\;\;\;\;\;\;\;\quad\quad\quad \phi_{n} \sim
\phi'_0+2\sqrt{3}n(1-q')\nonumber\\ N_n & \sim &
{4a_0^2q'^{2n}(1-q')\over \sqrt{3}C'},\;\;\;\;\; P^a_{n+1}\sim
2\sqrt{3}{C'\over a_n}\label{s2}.\ffa One remarkable result from
these solutions is that the lapse $N_n$ is determined at each
instant by the equation, contrasting to the continuum theory where
$N$ is a free Lagrangian multiplier and may be set to one
uniformly. Unfortunately we notice from (\ref{rr}) that unless the
factor $q$ is very close to one the continuum limit can not be
achieved as the ``time'' level $n$ approaches to infinity. In
particular the continuum solution can only be achieved with a
certain accuracy as we choose the desired $q$. Nevertheless, this
discrete solution may give rise to the classical solution given in
(\ref{cs}). To see this we need set the lapse function to unity as
$n$ approaches to infinity, otherwise the comparison is not
meaningful. Consider the solutions (\ref{s1}) and note that \f
\epsilon N\sim N_n \sim {4a_n(a_{n+1}-a_n)\over \sqrt{3}C},\ff we
take the continuum limit and set $N=1$, leading to \f a\dot{a}\sim
1,\ff whose solution is $a\sim t^{1/2}$. Here we have also defined
that $t=n\epsilon$ and $\dot{a}={a_{n+1}-a_n\over \epsilon}$. If
$q\sim e^{\delta}$ where $\delta$ is a small number, then we have
$\delta\sim {ln(t)\over 2n}$, yielding the classical solution
$\phi\sim\phi_0+\sqrt{3}ln(t)$. Of course in this sense $q$ is not
a strict constant any more but approaches to one as $n$ goes to
infinity.

The cosmological solution from big crunch to big bang can be
obtained by combining the solution (\ref{s2}) with $q'<1$ when
$n\leq 0$ and (\ref{s1})with $q>1$ when $n\geq 0$. To be single
valued at $n=0$, we have adjoint conditions $a_0=a'_0, C=C',
\phi_0=\phi'_0$ and $ q+q'=2$. But we note that the conjugate
momentum of $a_n$ has a jump from $\pi^{a'}_{0^-}=2\sqrt{3}{C\over
a_0}$ to $\pi^{a}_{0^+}=-2\sqrt{3}{C\over a_0}$. Physically it
means the universe contracts as the time level $n$ increases from
$-\infty$ until $n=0$ and then rebound at the minimum size of the
universe, turning into an expanding phase as $n>0$. In this
discretized version, we find the singularity as t approaches to
zero is always avoided unless we set $a_0=0$ by hand. But in
generic points $a_0$ should be determined by the evolution of
scale factor from $n<0$ to $n=0$, which in classical limit
corresponds to the evolution from $t=-\infty$ to $t=0$. As a
matter of fact, all these intuitive considerations might be
invalid as the gravitation effect becomes so strong that the
quantum theory of gravity should be taken into account. Then the
quantum states of the universe is the superpositions of such
states describing expanding and contracting universes.
Nevertheless based on this heuristic analysis we believe the
evolution of universe becomes more delicate due to the quantum
effect of gravity and rather than a genuine crunch occurs, it
might rebound once a minimum size of the scale factor is
saturated.

We realize once the quantization of discrete theory is concerned,
we need a probability explanation of such special initial
conditions, namely why the recurrence constant $q$ is close to one
but not else. In classical theory we are short of such strategy to
present a satisfying explanation. Based on above consideration, we
see the closest argument appearing in cosmological model from big
crunch to big bang is the requirement that $q+q'=2$. If $a_n$ is
always positive as it should be, then $q$ and $q'$ could be any
positive number in $(0,2)$, not directly leading to our expected
value $e^{\delta}$ and $e^{-\delta}$ respectively. One might
expect that the classical solution (\ref{rr}) is not stable unless
the $q$ approaches to unity, in other words $q=1$ might be thought
of as an attractor and such running of $q$ from any initial
constant to unity might be implemented or required at quantum
mechanical level. But following perturbative analysis shows it's
not the case. Consider the perturbation \fa a_n &=&
a^{(0)}_n+\delta a_n,\quad
\phi_n=\phi^{(0)}_n+\delta \phi_n,\nonumber\\
P^a_n &=& P^{a(0)}_n+\delta P^a_n,\quad P^{\phi}_n=P^{\phi
(0)}_n+\delta P^{\phi}_n \nonumber\\ N_n &=& N^{(0)}_n+\delta
N_n.\ffa Substituting into equations of motion (\ref{eom}) and
keeping the first order correction, we have \f \delta
a_{n+1}+{2q-8\over 3}\delta a_n +q^2\delta a_{n-1}=0.\ff On the
other hand, perturbations directly based on (\ref{rr}) leads to \f
\delta a_{n+1}-2q\delta a_n +q^2\delta a_{n-1}=0.\ff The
consistency requires that $q\sim 1$, which looks satisfying. But
another kind of reasonable perturbation is $a_n=a_0(q+\delta q)^n
$ such that $\delta a_{n+1}={n+1\over n}q\delta a_n$. It
immediately leads to the conclusion that solutions (\ref{s1}) and
(\ref{s2}) are classically stable except for $q=1$, at that point
$\delta a_n$ could be any value, which is opposite to what we
desire.

We hope this dilemma may be solved by further progress in the
quantum theory of gravity. But before that we want to present
another model where a good continuum limit can be achieved nearby
the singularity in next section.

\section{Chaplygin Gas}
Chaplygin gas as an exotic fluid has been greatly investigated in
FRW cosmological model\cite{Chaplygin,Gibbons}( also see
\cite{Gibbons} for more references). Since it behaves like a
cosmological constant at ${\it later}$ time of the universe,
namely contributing a positive energy density and negative
pressure, it is supposed to be responsible for the acceleration of
the universe, as indicated by recent observation\cite{Ia}. But at
early time of the universe it behaves just like a dust matter such
that the effective cosmological constant is zero. One interesting
feature of Chaplygin gas is that its lagrangian is quite similar
to the effective lagrangian of tachyon field in string theory,
which has been proposed by A. Sen\cite{Sen1}, even though the
exact expression is not clear yet. The lagrangian of both
Chaplygin gas and tachyon fields can be given as   \f {\cal L}=
{-1\over 2N}\left ( \dot{a}^2+{8\pi Ga^4N^2V\over
3}\sqrt{1-{\dot{T}^2\over N^2a^2}}\right ).\ff For Chaplygin gas,
the potential $V(T)$ is a constant, while for tachyon fields
$V(T)$ has a maximum at $T =0$ and exponentially decreases for
large $T$\footnote{In particular, in\cite{Sen2} Sen conjectured
that the tachyon field may not be physical matter, but viewed as
the time variable in quantum cosmology as $\phi$ goes large. This
interesting treatment has its advantages in quantum cosmology and
more implications in nonperturbative quantum gravity may be
further studied and discussed elsewhere.}.

In this section we consider the discrete theory of Chaplygin
cosmology. To compare our result with those in previous section,
we still adopt the conformal coordinate system \f
ds^2=a^2(t)[-N^2(t)dt^2+dx_i^2].\ff As a matter of fact, we are
free to redefine the lapse function,  \f N'(t):=N(t)a(t),\ff and
then turn to \f ds^2=-{N'}^2(t)dt^2+a^2(t)dx_i^2.\ff The equations
of motion and constraints have the following familiar form  \fa
&&{1\over aN'}{d\over dt}\left( {\dot{a}\over N'}\right)={8\pi
G\over 3}\frac{V(T)}{\sqrt{1-{\dot{T}^2\over {N'}^2}}}\left(
1-{3\dot{T}^2\over 2{N'}^2}\right)\nonumber\\ &&
\frac{\dot{a}^2}{a^2{N'}^2} = {8\pi G\over
3}\frac{V(T)}{\sqrt{1-{\dot{T}^2\over {N'}^2}}}\nonumber\\
&& {N'\over N'^2-\dot{T}^2}{d\over dt}\left( {\dot{T}\over
N'}\right)+3{\dot{a}\dot{T}\over a{N'}^2}+{1\over V}{dV\over
dT}=0.\ffa As is familiar, we call the first two equations as
Raychaudhuri and Friedman equations respectively in standard
cosmology and the third one is the equation of motion for tachyon
fields. Furthermore, as these three equations are not independent
we are free to set the lapse function $N'=1$.

Next we consider the canonical formalism of the theory.  Define
the conjugate momenta of scale factor $a$ and scalar field $T$ \fa
\pi_a &=& {\partial {\cal L}\over
\partial \dot{a}}={-\dot{a}\over N}\nonumber\\ \pi_{T} &=& {\partial
{\cal L}\over \partial \dot{T}}={4\pi G\over 3}{a^3V\dot{T}\over
\sqrt{N^2a^2-\dot{T}^2}},\ffa  then the Hamiltonian is given as
\fa {\cal H}&=&\pi_a\dot{a}+\pi_T\dot{T}-{\cal L}\nonumber\\
&=& {-1\over 2N}\left ( \dot{a}^2-{8\pi Ga^4N^2\over 3}{V\over
\sqrt{1-{\dot{T}^2\over N^2a^2}}}\right ).\ffa The discretized
version of the lagrangian is  \fa L(n,n+1)&=&
\pia(a_{n+1}-a_n)+\pit(T_{n+1}-T_n)\nonumber\\
&+& {N_n\over2}\left[ \pia^2- 2a_n\sqrt{\pit^2+({4\pi Ga_n^3V\over
3})^2}\right].\ffa The canonically conjugate momenta of these
variables at instant $n+1$ and $n$ are respectively given by \fa
P^a_{n+1}=\pia &&\;\;\;\;\; P_n^a=\pia+ N_n\left[ \lambda_n+({4\pi
GV\over 3})^2{3a_n^6\over \lambda_n}\right]\nonumber\\
P^{\pi_a}_{n+1}=0 &&\;\;\;\;\;
P_n^{\pi_a}=-\left[(a_{n+1}-a_n)+N_n\pia\right]\nonumber\\
P^{\pi_T}_{n+1}=0 &&\;\;\;\;\;
P_n^{\pi_T}=-\left[(T_{n+1}-T_n)-{N_na_n\pit\over \lambda_n}\right]\nonumber\\
P^T_{n+1}=\pit &&\;\;\;\;\; P^T_n=\pit + \left( {4\pi Ga_n^3\over
3}\right)^2{N_na_nV\over \lambda_n}{dV\over dT}\nonumber\\
P^N_{n+1}=0 &&\;\;\;\;\; P^N_n={1\over 2}(\pia^2-
2a_n\lambda_n),\ffa where we have defined  \f \lambda_n\equiv
\sqrt{\pit^2+\left( {4\pi Ga_n^3V\over 3}\right)^2}.\ff In
discrete theory the time evolution is described by a canonical
transformation from the variables from $(q_n, p_n)$ to $(q_n,
p_{n+1})$. Taking $(a_n, T_n, P^a_n, P^T_n)$ as the set of basic
variables, we rewrite these equations of motion as
\fa  && a_{n+1}-a_n=-N_n\pia\nonumber\\
    && P^a_{n+1}-P^a_n=- N_n\left[ \lambda_n+3({4\pi GV\over
    3})^2{a_n^6\over \lambda_n}\right]\nonumber\\
    && T_{n+1}-T_n={N_na_n\pit\over \lambda_n}\nonumber\\
    && P^T_{n+1}-P^T_n=- \left( {4\pi Ga_n^3\over
3}\right)^2{N_na_nV\over \lambda_n}{dV\over dT}\nonumber\\
&& (P^{a}_{n+1})^2- 2a_n\lambda_n=0\ffa

Above consideration is applicable to both Chaplygin gas and
tachyon field. From now on we only focus on the case of Chaplygin
gas which has a constant $V$. The immediate consequence from above
equation is that $P_{n+1}^T$ is conserved, which is nothing but
$\pit$. First we consider the expanding universe with very small
scale factor comparing with $\pit$, \f \pit\gg ({4\pi Ga_n^3V\over
3}),\ff then $\lambda_n\simeq \pit=P^T_{n+1}$ and the equations of
motion reduce to
\fa  && a_{n+1}-a_n=-N_n\pia\nonumber\\
    && P^a_{n+1}-P^a_n=- N_n\pit \nonumber\\
    && T_{n+1}-T_n=N_na_n\nonumber\\
    && P^T_{n+1}-P^T_n=0\nonumber\\
&& (P^a_{n+1})^2- 2a_nP^T_{n+1}=0.\label{deg}\ffa

After solving for $N_n$ and $P^a_n$ from the second and fifth
equation, we may obtain the recurrence relations of $a_n$ from the
first equation as \f
a_{n+1}^2-6a_{n+1}a_n+9a_n^2-4a_na_{n-1}=0.\label{da}\ff

As we mentioned the evolution scheme is given by the canonical
transformation from $n$ to $n+1$. Given the initial conditions,
the value of scale factor can be determined at any instant by the
recurrence relation (\ref{da}). In particular, whether the
universe expands or contracts depends only on these initial data.
For instance if we take $a_1>a_0$, then we will see $a_{n+1}>a_n$
for any $n>1$, implying an expanding universe. On the other side,
if we choose $a_{-1}>a_0$, then we find $a_{n+1}<a_n$ for any
$n<-1$, hence implying an contracting universe. Similar to the
discussion in previous section, the universe from big crunch to
big bang may be obtained by connecting these two solutions at
$n=0$. We may also plot graphs to illustrate the discrete
evolution of $a$ by choosing appropriate parameters  such that the
scale factor goes to zero as $n$ approaches to $0$ from both $n<0$
and $n>0$, implying a singularity at $t=0$ in the continuum limit.
But in discrete theory more detailed analysis indicates that scale
factor takes a small but non-zero value at $n=0$ in general. Thus
its solutions may also be used to describe a universe from big
crunch to big bang but avoiding the singularity.

We remark that it is a conjecture that the universe from big
crunch to big bang may be obtained by connecting two solutions at
singularity, but not a simple solution to classical
equations\cite{Khoury}. In our discrete theory it says that the
set of $(a_{-1},a_0,a_1)$ does not satisfy the classical equations
(\ref{rr}) or (\ref{da}). The dynamical explanation of how the
universe rebounds at the minimum size $a_0$ is absent in this
classical framework, but it is not unreasonable as we expect the
quantum theory of discrete gravity would play an essential role to
solve this issue.

Now we turn to consider the continuum limit as $n$ goes to
infinity. This can be done by recovering the first and second
order derivatives with respect to time through \fa &&
\dot{a}={a_{n+1}-a_n\over \epsilon}\nonumber
\\&& \ddot{a}={a_{n+1}-2a_n+a_{n-1}\over \epsilon^2}.\ffa Then
equation (\ref{da}) becomes \f \dot{a}^2-4a\ddot{a}=0.\ff The
general solution to this equation is \f a=c_1(t+c_2)^{4\over
3},\ff where $c_1$ and $c_2$ are two constants, viewed as a
remnant of the reparameterization invariance of the continuum
theory.  Now going back to the other equations in (\ref{deg}) and
taking the continuum limit we straightforwardly obtain the
asymptotical behavior of other parameters  \f \pia\sim t^{2/3}
\;\;\;\;\;\;\; N_n\sim t^{-1/3}\;\;\;\;\;\; T\sim t^2, \ff and
thus the metric has the form  \f ds^2\sim
t^{8/3}(-t^{-2/3}dt^2+dx_i^2).\ff To see Chaplygin gas may be
though of as a dust matter during this epoch we turn to the
following conventional coordinate system with lapse function
$N'=1$: \f ds^2\sim -dT^2+T^{4/3}dx_i^2\label{tu}.\ff Therefore
the ``effective'' scale factor behaves as $a'\sim T^{2/3}$,
indicating an expanding but decelerating universe as $T>0$. One
interesting point here is that during the origin of the universe
the Chaplygin gas field may be understood as the proper time of
the universe. Such kind of behavior is something like the time
reversal of tachyon field, as pointed out in \cite{Sen2}, where
tachyon may play a role of time variable as rolling down to
infinity at ${\it later}$ time of the universe.

To see our discrete scheme actually leads to the same physics as
the continuum field theory of Chaplygin gas, we may also consider
the continuum limit for large scale factor. In this case we have,
\f \pit\ll ({4\pi Ga^3V\over 3}).\ff Since $V$ is a constant, then
\f \lambda_n\simeq {4\pi Ga^3V\over 3},\ff and the equations
reduce to
\fa  && a_{n+1}-a_n=-N_n\pia\nonumber\\
    && P^a_{n+1}-P^a_n=- 4N_n\lambda_n \nonumber\\
    && T_{n+1}-T_n={N_na_n\pit\over \lambda_n}\nonumber\\
    && P^T_{n+1}-P^T_n=0\nonumber\\
&& (P^a_{n+1})^2- 2a_nP^T_{n+1}=0.\label{ec2}\ffa Similarly we
obtain the recurrence relation of $a_n$ \f
2a_na_{n+1}-3a_n^2+a_{n-1}^2=0.\ff Taking the continuum limit
yields the differential equation \f 2a\ddot{a}+\dot{a}^2=0.\ff One
solution to this equation is $a\sim t^{2/3}$. Substituting it back
to (\ref{ec2}) we obtain the asymptotical behavior of other
variables as follows \f \pia\sim t^{4/3} \;\;\;\; N_n\sim
t^{-5/3}\;\;\;\; T\sim t^{-2}.\ff As a result, the metric becomes
\f ds^2\sim -{1\over T^2}dT^2+T^{-2/3}dx_i^2.\ff We note that the
Chaplygin gas does not play a  role of proper time any more. In
effect in the coordinate system with $N'=1$ the metric goes to \f
ds^2\sim -d\tau^2+\exp{{4\over 3}\tau}dx_i^2.\ff It indicates the
universe is accelerating, thus with a positive cosmological
constant, which is consistent with the result of continuum theory of Chaplygin gas.

\section{Conclusions}
We have applied discrete gravity theory to investigate the
universes from big crunch to big bang, mainly concerning the
following two issues. One is on the fate of cosmological
singularity. We proposed an attractive point of view on this issue
based on the consistent discretization approach to general
relativity. On the other hand we investigated the continuum limit
of discrete theories describing the universe with vanishing
cosmological constant. It's clear one may take the attempts
presented here as the first step to understand what role the
discrete gravity would play in quantum gravity and cosmology. In
particular in this paper we leave all the considerations on the
quantization of discrete gravity for further investigation. We
expect the approach to discrete general relativity will show more
insight into the problem of the cosmological singularity and
quantum gravity.

\section*{Acknowledgement}
I would like to thank Jorge Pullin and Lee Smolin for
correspondence and helpful discussions. This work is partly
supported by NSFC (No.10205002), SRF for ROCS, SEM and K.C.Wong
Education Foundation, Hong Kong.

\end{document}